\documentclass[aip,reprint]{revtex4-1}
\usepackage[T1,T2A]{fontenc}
\usepackage[utf8]{inputenc}
\usepackage[russian,english]{babel}
\usepackage{graphicx}
\usepackage{dcolumn}
\usepackage{bm}
\usepackage{graphicx}
\usepackage{amsmath}
\usepackage{amssymb}
\usepackage{pifont}
\usepackage{epstopdf}
\usepackage{color}

\begin{document}

\title{Self-diffusiophoresis of Janus particles that release ions}
\author{Evgeny S. Asmolov (Евгений С. Асмолов)}
\affiliation{Frumkin Institute of Physical Chemistry and
Electrochemistry, Russian Academy of Science, 31 Leninsky Prospect,
119071 Moscow, Russia}
\affiliation{Institute of Mechanics, Lomonosov Moscow State
University, 1 Michurinskiy Prospect, 119991 Moscow, Russia}
\email[Corresponding author: ]{aes50@yandex.ru}

\author{Tatiana V. Nizkaya (Татьяна В. Низкая)}
\affiliation{Frumkin Institute of Physical Chemistry and
   Electrochemistry, Russian Academy of Science, 31 Leninsky Prospect,
   119071 Moscow, Russia}

\author{Olga I. Vinogradova (Ольга И. Виноградова)}
\affiliation{Frumkin Institute of Physical Chemistry and
   Electrochemistry, Russian Academy of Science, 31 Leninsky Prospect,
   119071 Moscow, Russia}

\begin{abstract}
Catalytic Janus swimmers demonstrate a diffusio-phoretic motion by self-generating the gradients of concentrations and electric potential. Recent work has focused on simplified cases, such as a release of solely one type of ions
or low surface fluxes of ions, with limited theoretical guidance. Here, we consider the experimentally relevant case of particles that release both types of ions, and obtain a simple expression for a particle velocity in the limit of thin electrostatic diffuse layer.  Our approximate expression is very accurate  even
when ion fluxes and surface potentials are large, and allows one to interpret a number of intriguing phenomena, such as the reverse in the direction of the particle motion in response to variations of the salt concentration or self-diffusiophoresis of uncharged particles.

\end{abstract}

\maketitle

\affiliation{Frumkin Institute of Physical Chemistry and
   Electrochemistry, Russian Academy of Science, 31 Leninsky Prospect,
   119071 Moscow, Russia}

\section{Introduction}

Catalytic swimmers have received a lot of attention in recent years \cite%
{dey2017,moran2017}. They serve as a model system to study collective
behavior of active particles \cite{buttinoni2013dynamical} and have
potential applications in drug delivery and nano-robotics \cite{hu2020micro}%
. The propulsion mechanism of Janus catalytic swimmers is associated with an
inhomogeneous production of species at the particle surface \cite%
{golestanian2005propulsion}. A part of the particle is chemically active and
catalyzes a reaction with the consequent a flux of the reaction
products from the surface to the surrounding solution. In contrast, another part of the particle is chemically inert or catalyzes a
reverse reaction. The resulting concentration gradient of products induces a directed migration of the particle relative to a fluid reffered to as a self-diffusiophoresis
\cite{golestanian2005propulsion,golestanian2007}. Since the reaction products often represent ions of different diffusion rates, they induce an electric field at distances comparable to the particle radius. Such a field slows down (speeds up) ions with greater (smaller) diffusion coefficients, that naturally impacts the self-electrophoretic velocity of particles. \cite%
{paxton2005motility,moran2010locomotion}

The most known examples of catalytic swimmers are bi-metallic particles in a hydrogen
peroxide solution\cite{paxton2004,wang2006bipolar}, which are known to release  solely one type
of ions (namely, H$^+$). However, rapid development of enzymatic motors\cite%
{hermanova2020biocatalytic} has raised interest in swimmers that can produce both anions and cations (NH$_4^+$, HCO$_3^-$ for
urease motors\cite{patino2018fundamental}, Ag$^{+}$ and Cl$^{-}$ for AgCl
particles\cite{ibele2009}). Despite at least a decade of intense research on this kind of
motors, a body of  theoretical publications remain rather scarce, and  their quantitative understanding  is still challenging \cite{zhou2018,deCorato2020}.


A theoretical description of catalytic swimming involves the solution of the
Nernst-Planck equations for the concentration of ionic species, the Poisson equation
for an electric potential generated by an inhomogeneous charge distribution, and the
Stokes equation describing the fluid flow with the appropriate boundary conditions.

A solution containing ions builds up a so-called electrostatic diffuse layer (EDL) close to the charged particle, where the surface charge is balanced by the cloud of counterions. The extension of the EDL is defined by the screening length of an electrolyte solution, which is typically below a hundrend of nm\cite{poon.w:2006}. Far from the particle we deal with an outer electroneutral region (i.e. a bulk electrolyte). For microparticles the EDL is thin compared to their radius.
In this situation the problem can be solved  analytically by using
the method of matched
asymptotic expansions
 \cite%
{prieve.dc:1984,anderson.jl:1989,golestanian2007}. An outer solution is constructed at distances of the order of the particle size and allows one to obtain the values of concentrations and an electric field at the outer edge of the inner region defined within the EDL. Their gradients determine a flow within EDL. Since the EDL
is much thinner than radius, it appears macroscopically that the outer liquid slips over the surface. For this reason, the outer velocity is often termed an apparent
diffusio-osmotic slip velocity.  The velocity of the particle itself then can be evaluated by integrating the local apparent slip velocity over the particle surface. Such an approach has been widely applied for
systems that release one type of ions only\cite{yariv2011electrokinetic,sabass2012nonlinear,
nourhani2015,ibrahim2017}.
Some analytical solutions are known for the case\cite%
{nourhani2015,ibrahim2017} when the Nernst-Planck equations can be linearized, which is justified provided the variations of concentrations due to surface
flux are low. However, the outer problem in the non-linear case has been previously solved  only numerically \cite{sabass2012nonlinear}.

Finally, we mention that the boundary conditions generally reflect the
kinetics of chemical reactions at the surface\cite{moran2010locomotion}.
 Basically, two  approaches to modeling chemical activity of particles have been used. In an approach termed kinetic-based the fluxes depend on the instant local concentrations of ions \cite%
{moran2011electrokinetic,yariv2011electrokinetic,sabass2012nonlinear,ibrahim2017}. Thus, a detailed information on chemical reaction rates is required. A simpler flux-based approach we will use here assumes that  the fluxes of the reaction products are prescribed\cite%
{moran2010locomotion,nourhani2015}. 

In this paper we present a  theory of diffusio-phoretic motion of a Janus particle that  simultaneously releases cations and anions. Assuming the fluxes of both types of ions at the surface are equal and that the EDL is thin, we obtain the analytical expressions for the concentrations of ions, electric potential, and velocity of the particle. The equations are highly non-linear and hold even when ion fluxes and surface potentials are quite large. We show that the non-linearity leads to a number of startling results, such as the reversal of the direction of particle motion that can be tuned by salt or the arising diffusio-phoresis of neutral particles.

Our paper is organized as follows. The model and governing equations are formulated in
Sec.~\ref{s2}. Their solution via matched asymptotic expansions is given in
Sec.~\ref{s3}. In Sec.~\ref{s4} we present the numerical results for
concentration and electric fields and for the particle velocity. Our conclusions
are summarized in Sec.~\ref{s5}. In Appendix A we prove the equivalence of
the reciprocal theorem and the approach based on apparent slip velocity at
the particle surface\cite{anderson.jl:1989} in estimating the particle
velocity.

\section{Model and governing equations}
\label{s2}

\begin{figure}[t]
\centering
\vspace{-0.5cm} \includegraphics[width=0.7\columnwidth]{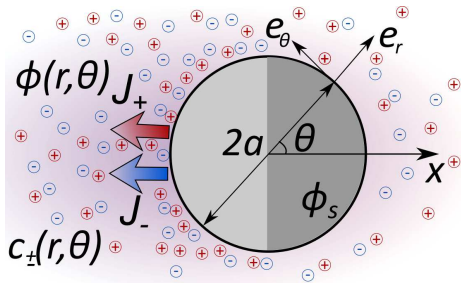} %
\caption{Sketch of a charged particle that releases ions from a portion of
its surface. }
\label{fig:sketch1}
\end{figure}

The basic assumptions of our model are as follows. We consider a charged spherical particle of radius $a$ immersed in a 1:1 bulk
electrolyte solution of permittivity $\epsilon $ and constant concentration $c_{\infty }$ as sketched in Fig.~\ref{fig:sketch1}. The dimensionless
(scaled by $k_{B}T/e,$ where $e$ is the elementary positive charge, $k_{B}$
is the Boltzmann constant, and $T$ is the temperature) electrostatic
potential $\phi _{s}$ of its surface is uniform, but a chemically active
portion of the particle releases monovalent cations and anions with a
surface flux $J$. We assume that the Peclet number is small, so that the convection of ions can
be neglected, and their fluxes satisfy dimensionless Nernst-Planck equations that describe the ion flux as function of both
concentration gradient and a migration term due to the electric field
\begin{equation}
\nabla \cdot \mathbf{J}_{\pm }=-D_{\pm }\left[ \Delta c_{\pm }\pm \nabla
\cdot \left( c_{\pm }\nabla \phi \right) \right] =0,  \label{NP1}
\end{equation}%
where the coordinates are scaled by $a$, $\phi $ is the dimensionless local
potential, $D_{\pm }$ is the diffusion coefficient of cations/anions, and $%
c_{\pm }$ is their local concentration. We further assume a release
of both ion species to be equal, $\mathbf{J}_{+}\cdot \mathbf{n}=\mathbf{J}_{-}\cdot
\mathbf{n}$.

In spherical coordinates the boundary conditions at $r=1$
take the form
\begin{equation}
\partial _{r}c_{\pm }\pm c_{\pm }\partial _{r}\phi =-\frac{Jj\left( \theta
\right) a}{D_{\pm }}.  \label{bc_1}
\end{equation}%
Here we introduced the function $j\left( \theta \right) \geq 0$ that determines a distribution
of the chemically active portions of the particle. This function is assumed here to be axisymmetric and
satisfying
\begin{equation}
\int_{0}^{\pi }j\left( \theta \right) \sin \theta d\theta =1.  \label{norm}
\end{equation}%
Since far from the surface we deal with the bulk electrolyte solution, the boundary conditions there are set as
\begin{equation}
r\rightarrow \infty :\ c_{+}=c_{-}=c_{\infty },\ \phi =0.  \label{bc_inf}
\end{equation}%

It is now convenient to introduce a characteristic concentration $c^{\ast }$ in
the vicinity of the particle
\begin{equation}
c^{\ast }=c_{\infty }(1+\mathrm{Da}),  \label{cw}
\end{equation}
where
\begin{equation}
\mathrm{Da}=\frac{Ja}{Dc_{\infty }},\quad D=\frac{2D_{+}D_{-}}{D_{+}+D_{-}}.
\label{cw2}
\end{equation}%
Here the Damk\"{o}hler number $\mathrm{Da}$ is characteristics of the excess of ions
near the particle. When $\mathrm{Da}\ll 1$, the surface flux is weak, and
the system is close to equilibrium, so that $c^{\ast }\simeq c_{\infty }$.
However, if $\mathrm{Da}\gg 1$, the concentration of the released ions
significantly exceeds the bulk one, and we deal with a highly
non-equilibrium system.

At any point $r$ the potential $\phi$ satisfies the Poisson equation
\begin{equation}
\Delta \phi =\lambda ^{-2}\frac{c_{+}-c_{-}}{2c^{\ast }},  \label{pois1}
\end{equation}%
where the new dimensionless parameter $\lambda =\Lambda a^{-1}$ represents
the ratio of the screening length $\Lambda=\left( 8\pi e^{2}c^{\ast
}/\epsilon k_{B}T\right) ^{-1/2}$ to the particle radius. Here $\Lambda$ is defined similarly to the Debye length of the bulk electrolyte, but we use $c^{\ast
}$ instead of $c_{\infty}$. Additional assumption is that we only consider
a thin EDL limit, $\lambda \ll 1.$

The (diffusio-osmotic) fluid flow satisfies the Stokes equations with an electrostatic body force
\cite%
{anderson.jl:1989}:%
\begin{equation}
\mathbf{\nabla \cdot v}=0\mathbf{,\quad }\Delta \mathbf{v}-\mathbf{\nabla }p=%
\mathbf{f},  \label{NS}
\end{equation}%
where $\mathbf{v}$ is the dimensionless (scaled by $\dfrac{\epsilon k_{B}^{2}T^{2}}{4\pi \eta e^{2}a}$) velocity, $p$ the dimensionless pressure and $\mathbf{f}=-\Delta \phi \mathbf{\nabla }\phi$ is the body force that drives the flow.

We apply the no-slip boundary condition at the particle surface, $\mathbf{v}=%
\mathbf{v}_{p}$, where $\mathbf{v}_{p}$ is still unknown particle velocity. Our results thus apply only to hydrophilic (poorly wetted) particles, but not to hydrophobic ones where the hydrodynamic slip is expected\cite{vinogradova.oi:1999,vinogradova.oi:2011}.

We stress that to determine the velocity of a freely
moving  in the $x-$direction particle it is not necessary to solve the Stokes equations \eqref{NS} since it
can be found by using the reciprocal
theorem \cite{teubner1982,masoud2019}
\begin{equation}
v_{p}=-\frac{1}{6\pi }\int_{V_{f}}\mathbf{f}\cdot \left( \mathbf{v}_{1}-%
\mathbf{e}_{x}\right) dV  \label{v_rec}
\end{equation}%
The integral in the latter equation is evaluated over the whole fluid volume $V_{f}$ and $\mathbf{v}%
_{1} $ represents the velocity field for the particle of the same radius that translates
with the velocity $\mathbf{e}_{x}$ in a stagnant fluid (Stokes solution):
\begin{equation}
\mathbf{v}_{1}=\left( \frac{3}{2r}-\frac{1}{2r^{3}}\right) \cos \theta
\mathbf{e}_{r}-\left( \frac{3}{4r}+\frac{1}{4r^{3}}\right) \sin \theta
\mathbf{e}_{\theta }.  \label{v_st}
\end{equation}%
Therefore, once the coupled equations \eqref{NP1} and \eqref{pois1} for the ion
concentrations and the electric field are solved, the velocity of the phoretic Janus swimmer can be obtained
by integrating
Eq.\eqref{v_rec}.


\begin{figure}[tbp]
\centering
\vspace{-0.5cm} \includegraphics[width=0.65%
\columnwidth]{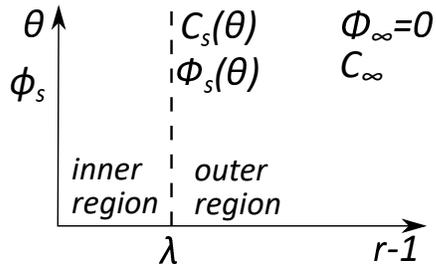}
\caption{Inner and outer regions for fields of concentration and electric potential.}
\label{fig:sketch2}
\end{figure}

\section{Theory}
\label{s3}

In this section we describe the theory of self-diffusiophoresis of a catalytic Janus particle. An accurate approximation to the solution to a system of governing equations described in Sec.~\ref{s2} is constructed by using the method of matched asymptotic expansions. The two domains of different length scales are schematically depicted in Fig.~\ref{fig:sketch2}. The length scale of the outer
region is the particle radius $a$, but for the inner region it is the screening
length $\Lambda$. For convenience we will below denote the outer dimensionless potential as $\Phi$, by keeping the former notation $\phi$ only for an inner domain.
We recall that the dimensionless parameter $\lambda$ defined above is considered to be small.

\subsection{Outer solution}

Let us denote by $C$ the dimensionless (scaled by $c^{\ast })$ concentration in the outer region and rewrite the Poisson
equation, Eq.\eqref{pois1}, as
\begin{equation}
\Delta \Phi =\lambda ^{-2}\frac{C_{+}-C_{-}}{2}.  \label{pois}
\end{equation}%
Since $\lambda \ll 1,$ the leading-order solution of \eqref{pois} is $%
C_{+}=C_{-}=C,$ i.e. the electroneutrality holds to $O\left( \lambda
^{2}\right) .$ However, one should take into account the small charge $%
C_{+}-C_{-}=O\left( \lambda ^{2}\right) $, since it induces a finite
potential difference in the outer region. Eqs.\eqref{NP1} then may be reexpressed as
\begin{eqnarray}
\Delta C+\nabla \cdot \left( C\nabla \Phi \right) &=&0,  \label{NP2} \\
\Delta C-\nabla \cdot \left( C\nabla \Phi \right) &=&0.  \label{NP3}
\end{eqnarray}%
Summing up and subtracting Eqs.\eqref{NP2} and \eqref{NP3}, and applying (\ref{bc_1}) we obtain
\begin{eqnarray}
\Delta C &=&0,  \label{dif} \\
\nabla \cdot \left( C\nabla \Phi \right) &=&0,  \label{pot2}
\end{eqnarray}%
with the boundary conditions at $r=1$:
\begin{gather}
 \partial _{r}C=-\left( 1-C_{\infty }\right) j\left( \theta \right) ,
\label{bc1} \\
C\partial _{r}\Phi =\beta \left( 1-C_{\infty }\right) j\left( \theta \right)
,  \label{bc_p}
\end{gather}%
where
\begin{equation}
\beta =\frac{D_{+}-D_{-}}{D_{+}+D_{-}},\ C_{\infty }=\frac{c_{\infty }}{
c_{\ast }}=\frac{1}{1+\mathrm{Da}}.  \label{bet}
\end{equation}%
We stress that the above dimensionless parameters are limited, namely $\left\vert \beta
\right\vert \leq 1$ and $0\leq C_{\infty }\leq 1$.

While Eqs.\eqref{dif} and \eqref{pot2} are identical to those derived before  for a Pt-insulator Janus swimmer\cite%
{nourhani2015,ibrahim2017}, when  cations only are released from the surface, the
boundary conditions \eqref{bc1} and \eqref{bc_p} are different. In the present case they correspond to a release of both types of ions with an equal surface flux.

The solution of Eq.\eqref{dif} with the imposed boundary conditions \eqref{bc1} can
be readily expressed in terms of Legendre polynomials,
\begin{equation}
C=C_{\infty }+(1-C_{\infty })\sum_{n=0}^{\infty }\frac{j_{n}}{n+1}%
P_{n}\left( \cos \theta \right) r^{-n-1},  \label{c_sol}
\end{equation}%
where
\begin{equation}
j_{n}=\left( n+1/2\right) \int_{0}^{\pi }j\left( \theta \right) P_{n}\left(
\cos \theta \right) \sin \theta d\theta .  \label{qn}
\end{equation}

The general solution of Eq.(\ref{pot2}) for the electric potential is still challenging. Simple expressions have been proposed before only for a situation
when the surface flux is
weak, $\mathrm{Da}\ll 1,$ and concentration disturbances are small\cite{nourhani2015,ibrahim2017}. In this case  $C\simeq 1$ and Eq.\eqref{pot2} can be linearized, by reducing to the Laplace equation for the potential. However, it is easy to verify that the function
\begin{equation}
\mathbf{\nabla }\Phi =-\frac{\beta }{C}\mathbf{\nabla }C=-\beta \mathbf{%
\nabla }\left( \ln C\right) .  \label{df_o}
\end{equation}%
satisfies non-linear \eqref{pot2} with boundary condition \eqref{bc_p}.
Equation (\ref{df_o}) is equivalent to the condition of zero flux, $\mathbf{J}_{+}-\mathbf{%
J}_{-}=\mathbf{0}$. Such a zero flux condition is traditionally applied for the
tangent component of the flux at the outer edge of the inner domain\cite{prieve.dc:1984}. We stress that our
Eq. (\ref{df_o}) remains applicable in the entire outer region. A corollary from this result is that besides the  potential gradient, it is possible to find
the potential itself by applying the boundary condition \eqref{bc_inf}
\begin{equation}
\Phi =-\beta \left[ \ln \left( C\right) -\ln \left( C_{\infty }\right) %
\right] .  \label{phi_o}
\end{equation}%
From Eq.\eqref{pois} it follows then that the concentration profile satisfies
\begin{equation*}
C_{+}-C_{-}=2\lambda ^{2}\Delta \Phi =-2\beta \lambda ^{2}\Delta \left( \ln
C\right) .
\end{equation*}

Equation (\ref{phi_o}) predicts that the outer potential is large when $C_{\infty }$ is small, which follows from the asymptotic behavior of $C$ and $\Phi $ at large distances, $r\gg 1$. Indeed, the leading terms of the solutions (\ref{c_sol}) and %
(\ref{df_o}) are
\begin{gather}
C\simeq C_{\infty }+\frac{(1-C_{\infty })j_{0}}{r},\\
\mathbf{\nabla }\Phi \simeq -\frac{\beta (1-C_{\infty })j_{0}}{r\left[
rC_{\infty }+(1-C_{\infty })j_{0}\right] }\mathbf{e}_{r}.  \label{far1}
\end{gather}%
Therefore, the electric field decays as $r^{-2}$.
However, if $C_{\infty }$ is small, such an asymptotic behavior could happen only at $r\gg C_{\infty }^{-1}$. In the situation when $1\ll
r\ll C_{\infty }^{-1}$, the field should decay as $r^{-1}$, i.e. slower. The latter asymptotic expression, when integrated over $r$, gives for the potential near the particle $\Phi\simeq \beta \ln
\left( C_{\infty }\right) \gg 1$. Thus, although the electric field is low, it is of very long range, and this range grows with a decrease of $C_\infty$.  Note, however, that the limiting case of vanishing $C_\infty$ cannot be  attained in steady state, due to a finite flux from the particle.

The outer solution provides us with the concentration $C_s$ and the potential $\Phi_s$ at the border with the inner region. In other words, these inner limits of the outer solution are to be used as
the outer limits for the inner problem. Mathematically  they can be formulated as
\begin{equation}
r\rightarrow 1+0:\
\begin{array}{ll}
\Phi =\Phi \left( 1,\theta \right) =\Phi _{s}\left( \theta \right) , &  \\
C=C\left( 1,\theta \right) =C_{s}\left( \theta \right) .&
\end{array}%
\label{in_l}
\end{equation}%

Thus, the outer solution affects the inner one due to gradients of both  concentration and potential induced at the border between regions. The
effect is, of course, small, when $\mathrm{Da}\ll 1$ and the problem can be linearized (see Sec. \ref{s_lin}). However, it becomes  significant in the non-linear case, especially when $\mathrm{Da}\gg 1.$ If so,
$C_{\infty }\ll 1$ and $\Phi _{s}\gg 1$ that follows from Eqs. \eqref{bet} and \eqref{phi_o}.

\subsection{Inner solution}

The inner solution can be constructed by using a stretched coordinate $\rho
=(r-1)/\lambda $. The dimensionless potential and concentrations in the
inner region are sought in the form $\phi = \Phi _{s}+\varphi \left( \rho ,\theta
\right) $ and $C=C_{s}\xi _{\pm }\left( \rho ,\theta \right) ,$ respectively.
The outer limits of the inner solution are obtained by matching with the
inner limits of the outer solution, Eq.\eqref{in_l},
\begin{equation}
\rho \rightarrow \infty :\quad \ \varphi =0,\ \xi _{\pm }=1.  \label{ou_l}
\end{equation}

The ion fluxes in terms of $\rho $ read $\lambda ^{-1}\left( C_{s}\partial _{\rho }\xi _{\pm }\pm C_{s}\xi _{\pm
}\partial _{\rho }\varphi \right),$ so they are of the order of $\lambda ^{-1}$ or much greater than the surface
fluxes $\frac{Jj\left( \theta \right) a}{D_{\pm }}=O\left( 1\right).$ Therefore, in the inner region
the latter can safely be neglected, and we can then consider that the ion concentration fields satisfy the Boltzmann distribution%
\begin{equation}
\xi _{\pm }=\exp \left( \mp \varphi \right) ,  \label{Bol}
\end{equation}%
and, consequently, that the potential obeys the Poisson-Boltzmann equation
\begin{equation}
\partial _{\rho \rho }\varphi =C_{s}\sinh \varphi .  \label{PB}
\end{equation}%
At the surface, $\rho =0$, the boundary condition reads
\begin{equation}
\varphi(0,\theta) =\phi _{s}-\Phi _{s}(\theta) \equiv \psi (\theta).  \label{phi_s}
\end{equation}
Here $\psi $ is the potential jump in the inner region.  As a side note, this is similar, but not fully identical to a so-called zeta, or electrokinetic potential, which for hydrophilic surfaces represents itself a potential jump in a whole EDL and is equal to $\phi_s$.

To calculate the velocity of the particle $v_p$ using the reciprocal theorem, we
have to take the integral in Eq.\eqref{v_rec} over the whole fluid volume that
includes both regions. The contributions of the outer and
the inner regions into the volume force in \eqref{v_rec} can be decomposed
\begin{equation}
\mathbf{f}=-\Delta \Phi \mathbf{\nabla }\Phi -\Delta \varphi \mathbf{\nabla }%
\left( \Phi +\varphi \right) -\Delta \Phi \mathbf{\nabla }\varphi .
\label{f_exp}
\end{equation}%
The first and second terms are associated with the outer and inner regions, correspondingly. We remark that $\Delta \Phi $ is small compared to $\Delta \varphi $ inside the EDL, while $\mathbf{\nabla }\varphi $ is negligible in the outer domain. Therefore, the last term in \eqref{f_exp} can safely be neglected in both regions and  the particle velocity represent a superimposition of two velocities,
\begin{equation}
v_{p}=v_{po}+v_{pi},  \label{v_full}
\end{equation}%
that correspond to the contributions of the outer and inner regions to the integral (\ref{v_rec}).

The first term in \eqref{f_exp} can
be rewritten as%
\begin{equation*}
\mathbf{f}_{o}=-\Delta \Phi \mathbf{\nabla }\Phi =-\beta ^{2}\frac{\left(
\mathbf{\nabla }C\right) ^{2}\mathbf{\nabla }C}{C^{3}},
\end{equation*}%
indicating that the contribution of the outer region to the particle velocity is quadratic in $\beta $
and cubic in concentration gradient:
\begin{equation}
v_{po}=-\frac{\beta ^{2}}{6\pi }\int_{V_{f}}\frac{\left( \mathbf{\nabla }%
C\right) ^{2}\mathbf{\nabla }C\cdot \left( \mathbf{v}_{1}-\mathbf{e}%
_{x}\right) }{C^{3}}dV.  \label{v_o}
\end{equation}%
Here the integral for a given $j\left( \theta \right) $ depends on $C_{\infty
} $ only. Later we shell see (see Sec.~\ref{s4})
that this contribution is small compared to that of the inner region.

As shown in \cite{anderson.jl:1989} the contribution to the particle velocity from the inner region can be found by integrating the diffusio-osmotic
slip velocity (defined at the outer edge of the EDL) over the surface,%
\begin{equation}
v_{pi}=-\frac{1}{S}\int_{S_{p}}\left( \mathbf{v}_{s}\cdot \mathbf{e}%
_{x}\right) dS=-\frac{1}{2}\int_{0}^{\pi }v_{s}\sin ^{2}\theta d\theta ,
\label{v_pedl}
\end{equation}%
where $S_{p}$ is the particle surface, and $\mathbf{v}_{s}$ is given by\cite{prieve.dc:1984}
\begin{equation}
\mathbf{v}_{s}={v}_{s}\mathbf{e}_{\theta }, \, {v}_{s}=\partial _{\theta }\left( \ln C_{s}\right) \left\{ -\beta
\psi +4\ln \left[ \cosh \left( \frac{\psi }{4}\right) \right] \right\},  \label{slip}
\end{equation}%
where $\psi$ is defined by Eq.\eqref{phi_s}. We recall that
$\psi$ can differ significantly from $\phi _{s}$ since $\Phi _{s}$ is
finite. Moreover, later we show that it can be quite large when $C_{\infty }$ is small (see Sec.~\ref{s4}).

The velocity $v_{pi}$ depends on three main dimensionless parameters, $%
\beta ,$ $C_{\infty }$ and $\phi _{s}.$ We can infer from (\ref{phi_o}), (%
\ref{v_pedl}) and (\ref{slip}) that the velocity remains the same when both $%
\beta $ and $\phi _{s}$ change the signs:%
\begin{equation}
v_{pi}\left( -\beta ,C_{\infty },-\phi _{s}\right) =v_{pi}\left( \beta
,C_{\infty },\phi _{s}\right) .  \label{sign}
\end{equation}%
%
%

The contribution of the inner region to the particle velocity can be also estimated in terms of the reciprocal theorem. By substituting \eqref{f_exp} to the volume integral (\ref{v_rec}) we obtain %
\begin{equation}
v_{pi}=-\frac{1}{6\pi }\int_{V_{f}}\Delta \varphi \left[ \mathbf{\nabla }%
\left( \Phi +\varphi \right) \cdot \left( \mathbf{v}_{1}-\mathbf{e}%
_{x}\right) \right] dV.  \label{v_rec_i}
\end{equation}%
In Appendix A we demonstrate that $v_{pi}$ given by (\ref{v_rec_i}) and the surface
integral (\ref{v_pedl}) are equal.



\subsection{Linearized solution at small surface flux\label{s_lin}}

We first consider the situation when the excess concentration due to ion
release is much smaller than the bulk concentration, $\mathrm{Da}\ll 1,\ $or
equivalently, $1-C_{\infty }\ll 1$, but $\phi _{s}$ is finite. In this case
we have $\left\vert \Phi _{s}/\phi _{s}\right\vert \ll 1$ and $\psi \simeq
\phi _{s}.$ Therefore, the linear approximation takes into account the
concentration gradient $\partial _{\theta }C_{s},$ but neglects the
variation of the potential $\Phi _{s}.$

The derivative in Eq. (\ref{slip}) can be approximated by
\begin{equation}
\partial _{\theta }\left( \ln C_{s}\right) \simeq \partial _{\theta
}C_{s}=-\left( 1-C_{\infty }\right) \sum_{n=0}\frac{j_{n}\sin \theta }{n+1}%
P_{n}^{\prime }\left( \cos \theta \right) .  \label{lin}
\end{equation}

After substituting (\ref{lin}) into Eq. (\ref{v_pedl}), we can conclude that only
the term with $n=1$ of Legendre expansion contributes to the integral, while
for $n\neq 1$ we have:
\begin{equation}
-\int_{0}^{\pi }P_{n}^{\prime }\left( \cos \theta \right) \sin ^{3}\theta
d\theta =\int_{-1}^{1}P_{n}P_{1}dx=0.  \label{leg}
\end{equation}

Contribution of the outer region to the integral (\ref{v_rec}) is $O\left(
(1-C_{\infty })^{3}\right) $ (see Eq. (\ref{v_o})) and can be neglected.
By using (\ref{lin}) we can then obtain
\begin{equation}
v_{p}=\left( 1-C_{\infty }\right) \left\{ \phi _{s}\beta -4\ln \left[ \cosh
\left( \frac{\phi _{s}}{4}\right) \right] \right\} \frac{j_{1}}{3}.
\label{v_lin}
\end{equation}%
Below we show that this linear solution represents a sensible approximation of
the exact results up to $1-C_{\infty }\simeq 0.5.$

%
%

\section{Results and discussion}
\label{s4}

\begin{figure}[t]
\centering
\vspace{-0.25cm} \includegraphics[width=1.0\columnwidth]{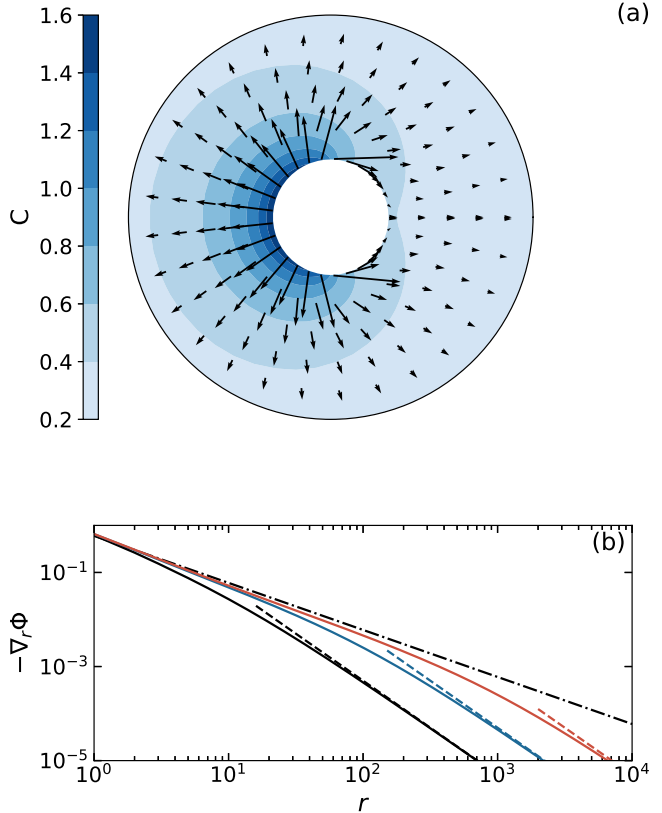}
\caption{(a) Distribution of concentration $C$ (color) and electric field $-%
\protect\nabla \Phi$ (arrows) in the outer region calculated using $\protect\beta=0.5$ and $C_\infty=0.01$. (b) The radial component of the electric field (solid curves) at $\theta=\pi/2$ as a function of radial coordinate. From left to right $C_\infty=0.1$, $0.01$, and $0.001$. Dashed lines show  the far-field asymptotes  obtained from (\ref{far1}). The dash-dotted line corresponds to the  $r^{-1}$ scaling.} 
\label{fig:fields}
\end{figure}

To illustrate the predictions of the general theory, we analyze a Janus particle with a piecewise constant distribution of the
flux in Eq.\eqref{bc_1}:
\begin{equation}
q(\theta )=\left\{
\begin{array}{ll}
0, & \theta \leq \pi /2, \\
2, & \pi /2<\theta \leq \pi .%
\end{array}%
\right.  \label{q_pw}
\end{equation}%
We first evaluate the concentration field in the outer region and use Eqs.\ (%
\ref{c_sol}), (\ref{qn}) truncating the sum at $N=34$, which approximates
the surface flux distribution (\ref{q_pw}) with a sufficient accuracy. Then
the electric potential $\Phi $ and the electric field $\mathbf{E}=-\nabla
\Phi $ are calculated using Eqs.\eqref{phi_o} and \eqref{df_o}.

\begin{figure}[b]
\centering
\vspace{-0.25cm} \includegraphics[width=1.0\columnwidth]{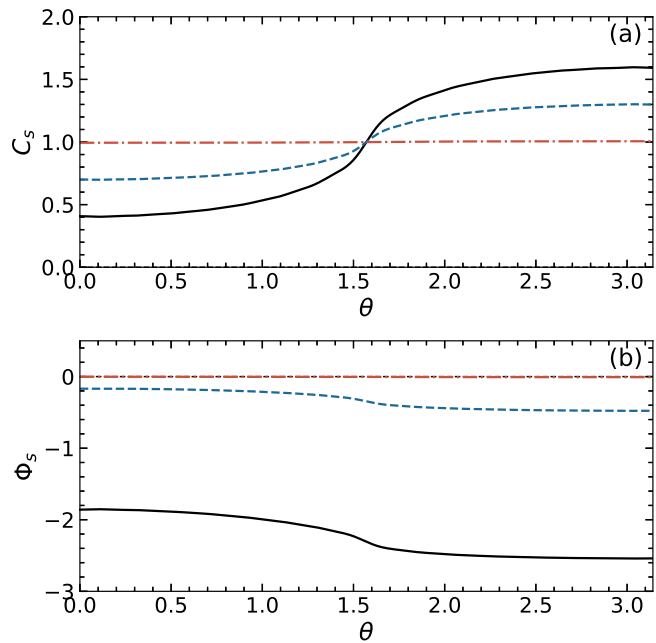}
\caption{(a) Ion concentration $C_s$ and (b) electric potential $\Phi_s$ 
as a function of $\theta$ calculated using $\protect\beta=0.5$ and $C_\infty=0.01$
(solid curve), $0.5$ (dashed curve) $0.99$ (dash-dotted curve). }
\label{fig:C_Phi}
\end{figure}

Figure \ref{fig:fields}(a) shows the distribution of the concentration and the
electric field in the outer region. The calculations are made for a low bulk concentration $C_{\infty
}=0.01$ and $\beta =0.5$. We see that the inhomogeneous surface flux (\ref{q_pw}) generates
large gradients of concentration and electric field near the left
(chemically active) side of the Janus particle both in the normal and tangent
directions.

The decay of the electric
field $\mathbf{E}$ in a far-field region, $r\gg
1$, is plotted in Fig. \ref{fig:fields}(b) in a log-log scale. For these numerical examples several concentrations of the bulk electrolyte are used. Also included are asymptotic results calculated from Eq.\eqref{far1} and a line corresponding to an inversely proportional to $r$ decay of the electric field.
It can be seen that for a largest concentration used, $C_{\infty }=0.1$,  the electric field decreases as $r^{-1}$ at $r=O(1)$ and begins to decay as $r^{-2}$ already at $r>50$.  The decrease in bulk concentration has the effect of slower decay of $\mathbf{E}$. If $C_{\infty }=0.001$, the $r^{-1}$ branch extends to $r=O(100)$  and a quadratic-law decay is observed only at distances that are nearly four order of magnitudes larger than the particle radius.


The angular distributions of concentration $C_{s}(\theta )$ and electric
potential $\Phi _{s}(\theta )$, obtained for several $C_{\infty }$, are shown in Fig. \ref{fig:C_Phi}. We see that in the limit $C_{\infty }\to 1$ (small
surface flux),  $C_s \simeq C_{\infty }$ and at any $\theta$ the potential $\Phi_s$ practically vanishes. On reducing $C_{\infty }$, the function $C_{s}(\theta )-1$ increases with $\theta$ and becomes antisymmetric. Its variations are greater at smaller $C_{\infty }$. However, 
the average concentration is always equal to unity due to our choice
of $c^{\ast }$. The potential $\Phi_s$ monotonously decreases with $\theta$, and its average value of the potential depends on
$C_{\infty }$,
\begin{equation}
\overline{\Phi _{s}}=\int_{0}^{\pi }\Phi _{s}(\theta )\sin \theta d\theta
\simeq \beta \ln (C_{\infty }),  \label{f_sa}
\end{equation}%
i.e. grows logarithmically with $C_{\infty }$ when the latter is sufficiently small. The function $\Phi
_{s}(\theta )-\beta \ln (C_{\infty })$ is also nearly antisymmetric.

\begin{figure}[t]
\centering
\includegraphics[width=1.01\columnwidth]{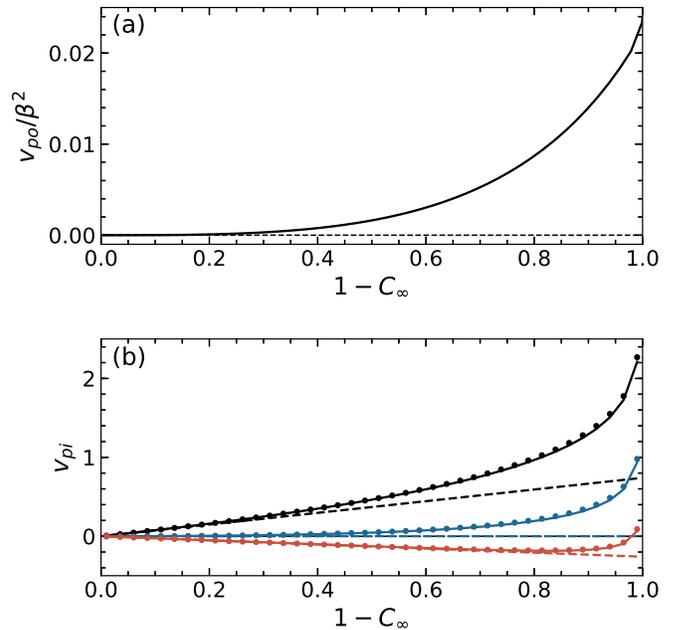}
\caption{(a) The velocity $v_{po}$, normalized by $\protect\beta ^{2}$, calculated from Eq. \eqref{v_o} and plotted against $1-C_{\infty }$. (b) The velocity $v_{pi}$ vs. $1-C_{\infty }$ calculated using $\protect\beta =0.5$ and $%
\protect\phi _{s}=2.5$, $0$, $-2.5$ (from top to bottom). Solid curves correspond
to the exact solution, Eq.\eqref{v_pedl}, dashed lines are calculated from
solution Eq.\eqref{v_lin}, circles show calculations from Eq.\eqref{fit}. }
\label{fig:vo_vi}
\end{figure}

\begin{figure}[t]
\centering
\includegraphics[width=1\columnwidth]{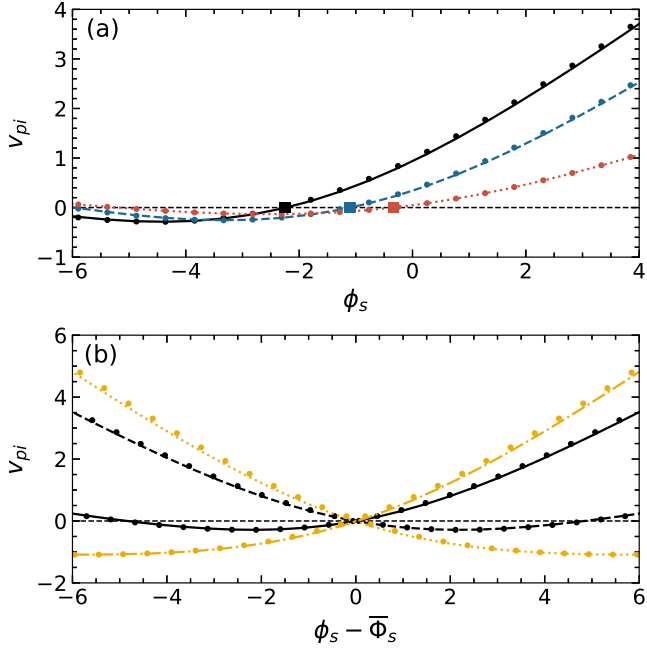}
\caption{(a) The velocity $v_{pi}$ as a function of $\phi_s$ calculated using $%
\protect\beta =0.5$ and $C_{\infty }=0.01$, $0.1$, and $0.5$ (solid, dashed, and dotted curves, correspondingly).
The large squares correspond to $\protect\phi %
_{s}=\overline{\Phi _{s}}$, circles show calculations from Eq.\eqref{fit}. (b) The velocity $v_{pi}$ plotted as a function of $\protect\phi %
_{s}-\overline{\Phi _{s}}$. Calculations are made for $C_{\infty }=0.01$ using $\protect\beta =-0.9,$ $-0.5,$ $0.5,$ and $0.9$ (dotted, dashed,
solid, dash-doted curves, correspondingly).   }
\label{fig:v_phi}
\end{figure}

\begin{figure}[tbp]
\centering
\includegraphics[width=1.0\columnwidth]{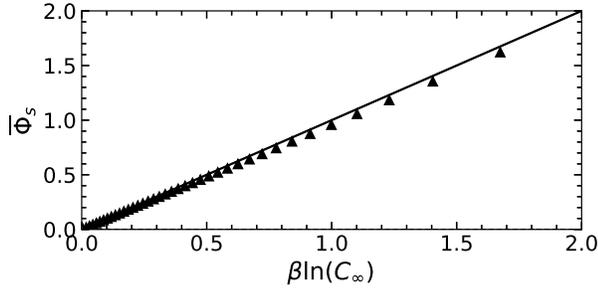}
\caption{Average electric potential at the border of the inner region vs. $\beta \ln C_{\infty }$ calculated using $\beta=0.5$. Triangles
correspond to $\overline{\Phi _{s}}$ calculated numerically, the solid line are calculations from Eq.(\protect\ref{f_sa}).}
\label{fig:Phi_lnC}
\end{figure}

Once the outer distributions of the ion concentration and the electric
potential are determined, we can evaluate the particle velocity using Eqs.(%
\ref{v_full}), (\ref{v_o}), and (\ref{v_pedl}). The surface integral Eq.%
\eqref{v_pedl} is calculated by applying a trapezoid rule on a uniform grid
in $\theta $ with $N_{\theta }=500$ nodes. To calculate the volume integral,
Eq. \eqref{v_o}, we use the same grid in $\theta $ and a non-uniform grid in $%
R$ (with a grid step varying as $R^{2}$) with $N_{r}=100$ nodes and a cut-off
radius $R_{out}=100$. 

The dependence of $v_{po}$ (scaled by $\beta
^{2}$) on $1-C_{\infty }$ is illustrated in Fig.~\ref{fig:vo_vi}(a). The magnitude of this scaled velocity grows on increasing $1-C_{\infty }$, but remains extremely small (it is well seen that $v_{po}/\beta
^{2}$ is below $0.025$ at $C_{\infty }=0$). 

The magnitude of 
$v_{pi}$  is much larger (see Fig.~\ref{fig:vo_vi}(b)). If 
$\beta $ and $\phi _{s}$ are positive, the velocity grows upon increasing $\phi _{s}$ and $1-C_{\infty }.$ One striking result is that even an
uncharged particle ($\phi _{s}=0$) can induce $v_{pi}=O(1)$ provided $C_{\infty }$ is small. Clearly, this happens due to a finite $\Phi _{s}$. Another important result is that in the case of negative
$\phi_s$, the velocity $v_{pi}$ can reverse its sign in response to the bulk electrolyte concentration. This phenomenon has been found before only numerically \cite{deCorato2020}, by using finite $\lambda$. Also included in Fig.~\ref{fig:vo_vi} are calculations from Eq.(\ref{v_lin}), obtained within a linear theory. The fits of numerical results are
quite good for relatively small $1-C_{\infty }$, but Eq.(\ref{v_lin}) significantly underestimates the results obtained at small $C_{\infty }$. 
It is also important to emphasise that  the reversal of the direction of the particle motion cannot be predicted using  a
linear theory.

If we keep $\beta =0.5$ fixed, but vary $\phi_s$ in a large range, we move to a situation shown in Fig.\ref{fig:v_phi}(a). These calculations are made using three different concentrations $C_{\infty }$, and we have marked with large squares the points that correspond to $\phi _{s} = \overline{\Phi _{s}}$. 
When $\phi _{s}$ is negative, the velocity $v_{pi}$ is rather small and reverse its sign at 
$\phi _{s}\simeq \overline{\Phi _{s}}$. Thus, it is clear that the
direction of the particle motion is controlled by the difference $\phi _{s}-\overline{%
\Phi _{s}}.$ In Fig. \ref{fig:v_phi}(b) we plot $v_{pi}$ as a function of $\phi _{s}-\overline{%
\Phi _{s}}$. These results are obtained for $C_{\infty }=0.01$ and $\beta =\pm 0.5,\ \pm 0.9$. Overall, these calculations confirm that the velocity reversal occurs when becomes equal to a surface potential. They also illustrate well  Eq. (\ref{sign}).

Finally, we check the validity of approximate Eq.~\eqref{f_sa}. Figure \ref{fig:Phi_lnC} plots $\overline{\Phi _{s}}$ as a function of $\beta \ln C_{\infty }$. We see that the approximation  $\overline{\Phi _{s}}\simeq \beta \ln C_{\infty }$ is quite accurate in a very
wide range of $C_{\infty }$. This reflects the fact that the variations of $\Phi _{s}$ are much smaller than
its average value which grows logarithmically as $C_{\infty }\to 0$.

We can thus suggest the following approximate formula for the diffusio-phoretic velocity of the catalytic Janus particle
\begin{equation}
v_{p}=\left\{ \frac{\beta \psi_{s}}{2}-2\ln \left[ \cosh \left( \frac{%
\psi_{s}}{4}\right) \right] \right\} \int_{0}^{\pi }\partial _{\theta
}\left( \ln C_{s}\right) \sin ^{2}\theta d\theta .  \label{fit}
\end{equation}%
where $\psi _{s}=\phi _{s}-\beta \ln (C_{\infty })$ and the integral can be readily
evaluated by using Eqs.\ (\ref{c_sol}) and (\ref{qn}). This equation takes into
account both variations of the concentration gradient and the electric field
along the particle surface. The calculations from Eq.\eqref{fit} are included into Figs. \ref%
{fig:vo_vi} and \ref{fig:v_phi}. A general conclusion from this
plot is that approximate equation \eqref{fit} is in excellent agreement with the numerical results, confirming the
validity of our theoretical description.

\section{Conclusion}

\label{s5}

We proposed a theory of a self-diffusiophoresis of the Janus
particle that release ions. The theory is valid in the limit of a thin electrostatic diffuse layer and employs the method of matched asymptotic expansions. We derived the analytic solution for both the concentration and the electric
field in the outer region. Our analysis and numerical calculations show that while the direct contribution of the outer region
to the particle velocity can be neglected, an inhomogeneous ion distribution
generates a finite potential at the outer edge of the inner region. As a result, even uncharged particles can migrate with a
finite velocity. We also showed that particles can reverse the direction of motion in response to
changes in bulk concentration of ions. Finally, we proposed a simple formula for a
particle velocity, which was found to be in excellent agreement with the numerical solution of the asymptotic
equations.

\begin{acknowledgments}

This work was supported by the Ministry of Science and Higher Education of the Russian Federation.
\end{acknowledgments}

\section*{DATA AVAILABILITY}

The data that support the findings of this study are available within the
article.

\appendix
\begin{widetext}

\section{Equivalence of the reciprocal theorem and the slip-velocity formula
in thin EDL}\label{ap}

We evaluate the contribution of the inner region to the particle velocity, i.e. calculate the integral (\ref{v_rec_i}). The auxiliary velocity field (\ref{v_st})
can be presented in terms of stretched variable $\rho $ as%
\begin{equation*}
\mathbf{v}_{1}-\mathbf{e}_{x}\simeq -\frac{3}{2}\rho ^{2}\lambda
^{2}\cos \theta \mathbf{e}_{r}+\frac{3}{2}\rho \lambda \sin \theta
\mathbf{e}_{\theta },
\end{equation*}%
i.e. the velocity is small in the inner region. However, the derivatives
of $\varphi $ with respect to $\rho$ are large: $\Delta \varphi \simeq \lambda ^{-2}\partial
_{\rho \rho }\varphi ,\ \mathbf{\nabla }\varphi =\lambda ^{-1}\partial _{\rho
}\varphi \mathbf{e}_{r}+\partial _{\theta }\varphi \mathbf{e}_{\theta }.$ As a
result, the integral (\ref{v_rec_i}) is finite:%
\begin{equation}
v_{pi}=-\frac{1}{6\pi }\int_{V_{EDL}}\mathbf{f}_{i}\cdot \left( \mathbf{v}
_{1}-\mathbf{e}_{x}\right) dV\simeq -\frac{1}{2}\int\limits_{0}^{\pi
}\int\limits_{0}^{\infty }\partial _{\rho \rho }\varphi \left[ \partial _{\rho
}\varphi \rho ^{2}\cos \theta -\partial _{\theta }\left( \Phi _{s}+\varphi \right)
\rho \sin \theta \right] d\rho \sin \theta d\theta .  \label{vfi2}
\end{equation}

We integrate the first term in the square brackets over $\rho $ using
integration by parts:%
\begin{equation}
-\frac{\sin 2\theta }{2}\int\limits_{0}^{\infty }\partial _{\rho \rho }\varphi
\partial _{\rho }\varphi \rho ^{2}d\rho =-\frac{\sin 2\theta }{2}%
\int\limits_{0}^{\infty }\rho ^{2}d\left[ \left( \partial _{\rho }\varphi
/2\right) ^{2}\right] =C_{s}\sin 2\theta \int\limits_{0}^{\infty }\left(
\cosh \varphi -1\right) \rho d\rho .  \label{f1ir}
\end{equation}%
The term proportional to $\partial _{\theta
}\Phi _{s}$ in (\ref{vfi2}) integrated over $\rho $ gives the usual electrophoretic slip velocity,%
\begin{equation}
\partial _{\theta }\Phi _{s}\int\limits_{0}^{\infty }\rho \partial _{\rho
\rho }\varphi d\rho =\partial _{\theta }\Phi _{s}\int\limits_{0}^{\infty }\rho
d\partial _{\rho }\varphi =-\partial _{\theta }\Phi _{s}\left( \phi
_{s}-\Phi  _{s}\right) .  \label{f2r}
\end{equation}%
To calculate the term proportional to $\partial _{\theta }\varphi $, we first integrate over $%
\theta $ using Poisson-Boltzmann equation (\ref{PB}) and integration by
parts:%
\begin{equation}
\rho \int\limits_{0}^{\pi }C_{s}\sinh \varphi \partial _{\theta }\varphi \sin
^{2}\theta d\theta =-\rho \int\limits_{0}^{\pi }\left( \cosh \varphi -1\right)
\sin 2\theta C_{s}d\theta -\rho \int\limits_{0}^{\pi }\left( \cosh \varphi
-1\right) \sin ^{2}\theta \partial _{\theta }C_{s}d\theta .  \label{f2t2}
\end{equation}%
Then the integral of the first term in (\ref{f2t2}) over $\rho $ and the
integral of (\ref{f1ir}) over $\theta $ cancel out. In the second term in (%
\ref{f2t2}), $\rho \left( \cosh \varphi -1\right) $ only depends on $\rho .$
To calculate its integral over $\rho $ we use the equality $\left( \partial
_{\rho }\varphi \right) ^{2}=2C_{s}\left( \cosh \varphi -1\right) $, which follows
from Eq. (\ref{PB}), and the variable change $d\rho =d\varphi /\sqrt{%
2C_{s}\left( \cosh \varphi -1\right) }=d\varphi /\left( \sqrt{C_{s}}\sinh \left(
\varphi /2\right) \right) :$%
\begin{align}
\int\limits_{0}^{\infty }\rho \left( \cosh \varphi -1\right) d\rho &
=\int_{0}^{\infty }\rho \frac{\sinh \left( \varphi /2\right) }{\sqrt{C_{s}}}%
d\varphi =-\int_{0}^{\infty }\frac{2}{\sqrt{C_{s}}}\left( \cosh \left( \varphi
/2\right) -1\right) d\rho =-\int_{0}^{\infty }\frac{\cosh \left( \varphi
/2\right) -1}{C_{s}\sinh \left( \varphi /2\right) }d\varphi  \\
& =-\int_{0}^{\infty }\frac{\sinh \left( \varphi /4\right) }{C_{s}\cosh \left(
\varphi /4\right) }d\varphi =\frac{4}{C_{s}}\ln \cosh \left( \zeta /4\right).
\label{f3r}
\end{align}

Finally, collecting all the terms from Eqs. (\ref{vfi2}), (\ref{f2r}), (\ref{f2t2}) and %
(\ref{f3r}), we obtain the following formula for the particle velocity:%
\begin{equation}
v_{pi}=-\frac{1}{2}\int\limits_{0}^{\pi }\left[ \psi \partial _{\theta }\Phi
_{s}+4\partial _{\theta }\left( \ln C_{s}\right) \ln \cosh \left( \psi
/4\right) \right] \sin ^{2}\theta d\theta ,
\end{equation}%
with \begin{equation*}
\psi (\theta )=\phi _{s}-\Phi _{s}(\theta ),
\end{equation*}%
which is fully equivalent to (\ref{v_pedl}).
\end{widetext}

\end{document}